\documentclass[twocolumn,english,amsart,showpacs,preprintnumbers,amsmath,amssymb,floatfix]{revtex4-1}
\usepackage{tikz,xcolor}
\usepackage[colorlinks = true,  
linkcolor = blue,
urlcolor  = blue,
citecolor = blue,
anchorcolor = blue]{hyperref} 

\definecolor{lime}{HTML}{A6CE39}
\DeclareRobustCommand{\orcidicon}{%
	\begin{tikzpicture}
	\draw[lime, fill=lime] (0,0) 
	circle [radius=0.16] 
	node[white] {{\fontfamily{qag}\selectfont \tiny ID}};
	\draw[white, fill=white] (-0.0625,0.095) 
	circle [radius=0.007];
	\end{tikzpicture}
	\hspace{-2mm}
}

\foreach \x in {A, ..., Z}{%
\expandafter\xdef\csname orcid\x\endcsname{\noexpand\href{https://orcid.org/\csname orcidauthor\x\endcsname}{\noexpand\orcidicon}}
}

\usepackage[T1]{fontenc}
\usepackage[latin9]{inputenc}
\usepackage{color}
\usepackage{array}
\usepackage{amstext}
\usepackage{graphicx}
\usepackage{esint}
\usepackage{rotating}
\usepackage{appendix}
\usepackage{float}
\usepackage{multirow}

\usepackage[font=small,labelfont=bf]{caption}
\usepackage{xcolor}
\usepackage{ulem}

\makeatletter

\@ifundefined{textcolor}{}
{%
 \definecolor{BLACK}{gray}{0}
 \definecolor{WHITE}{gray}{1}
 \definecolor{RED}{rgb}{1,0,0}
 \definecolor{GREEN}{rgb}{0,1,0}
 \definecolor{BLUE}{rgb}{0,0,1}
 \definecolor{CYAN}{cmyk}{1,0,0,0}
 \definecolor{MAGENTA}{cmyk}{0,1,0,0}
 \definecolor{YELLOW}{cmyk}{0,0,1,0}
 }

\@ifundefined{definecolor}
 {\usepackage{color}}{}
\@ifundefined{definecolor}
 {\usepackage{color}}{}
\makeatother
\usepackage{babel}

\newcommand{\delphes}{{\sc Delphes }}
\newcommand{\madgraph}{{\sc MadGraph }}

\newcommand{\feynrules}{{\sc Feyn\-Rules }}
\newcommand{\pythia}{{\sc Pythia }}

\newcommand{\beq}{\begin{equation}}
	\newcommand{\eeq}{\end{equation}}

\newcommand{\cL}{\mathcal{L}}

\def\Re{{\cal R \mskip-4mu \lower.1ex \hbox{\it e}\,}}
\def\Im{{\cal I \mskip-5mu \lower.1ex \hbox{\it m}\,}}

\def\tev{\,{\ifmmode\mathrm {TeV}\else TeV\fi}}
\def\gev{\,{\ifmmode\mathrm {GeV}\else GeV\fi}}
\def\mev{\,{\ifmmode\mathrm {MeV}\else MeV\fi}}

\begin{document}


\title{ Probing four-fermion operators in the triple top production at future hadron colliders  }

\author{Sara Khatibi$^{1}$\orcidA{}}
\email{Sara.Khatibi@ut.ac.ir}

\author{Hamzeh Khanpour$^{2,3,4}$\orcidB{}}
\email{Hamzeh.Khanpour@cern.ch}

\affiliation {
$^{(1)}$Department of Physics, University of Tehran, North Karegar Ave., Tehran 14395-547, Iran   \\ 
$^{(2)}$Department of Physics, University of Science and Technology of Mazandaran, P.O.Box 48518-78195, Behshahr, Iran    \\
$^{(3)}$School of Particles and Accelerators, Institute for Research in Fundamental Sciences (IPM), P.O.Box 19395-5531, Tehran, Iran\\
$^{(4)}$ Department of Theoretical Physics, Maynooth University, Maynooth, Co. Kildare, Ireland
}


%
\begin{abstract}\label{abstract}

In this paper, we study the triple top quark production at the 
future high-energy proton-proton colliders 
to probe the four-fermion interactions involving three top quarks.
We employ the Standard Model Effective Field Theory (SMEFT) 
to find the upper limits at $95\%$ CL on the Wilson coefficients 
of these kinds of four-fermion operators.
We consider a detailed analysis with a unique signal signature of 
two same-sign leptons.
A full simulation chain includes all the relevant backgrounds, 
realistic detector simulations, and a cut-based technique are 
taken into account. 
This study is presented for the HE-LHC working at the center 
of mass energy of 27 TeV with 15~ab$^{-1}$ and FCC-hh working at the center 
of mass energy of 100 TeV with 30~ab$^{-1}$.
We show that the future high-energy proton-proton colliders
could reach an impressive sensitivity to four-fermion contact 
interactions involving three top quarks.

\end{abstract}   
%



\maketitle


\section{Introduction} \label{intro}

All experimental measurements agree well with the 
Standard Model (SM) forecasts so far.
It suggests that the scale of new physics is much above the electroweak scale 
and these two scales are well separated from each other. In other words, 
the new heavy degrees of freedom cannot be produced directly in the experiments, 
however, they can affect the SM couplings or induce the new
interaction between the SM degrees of freedom.

The Standard Model Effective Field Theory (SMEFT) provides a 
framework for studying the effect of new physics~\cite{Brivio:2017vri}. 
The SMEFT Lagrangian only contains the SM particles and respects the 
Lorentz and the SM gauge symmetry, i.e. SU(3)$\times$SU(2)$\times$U(1).
In this framework, the new degrees of freedom are integrated out and 
their effects show up as the new interaction between the SM particles 
in the form of higher dimension operators in the Lagrangian. 
The higher dimensional terms are proportional to the inverse powers of 
the scale of new physics. Assuming baryon number conservation, the leading 
corrections to the SM interactions come from one dimension-five and 59 
dimension-six independent operators~\cite{Grzadkowski:2010es}.

Being a model-independent approach is one of the advantages of the SMEFT framework
since the various new physics scenarios exist in the market which sometimes 
have the same signatures in the experiment environments. Hence, one can use this 
model-independent framework and find the experimental values or limits 
on the Wilson coefficients of higher-dimensional operators and then translate 
them to the matching couplings of any new physics model. 
Among dimension-6 operators, four-fermion contact interaction 
can be originated at tree level in the UV models so they can have 
large Wilson coefficients~\cite{deBlas:2017xtg}.
There are several papers
in the literature that has studied the four-fermion 
interactions using different observables. In Ref.~\cite{Domenech:2012ai}, 
the dijet measurements at the Large Hadron Collider (LHC) 
have been utilized to search for 
four-light quark operators.
In addition, the four-lepton contact interactions have 
been considered in Ref.~\cite{delAguila:2014soa,Falkowski:2015krw}.

Due to the interesting properties of top quark, studying the 
four-fermion operators involving the top quark is also important. 
Ref.~\cite{DHondt:2018cww} constrained the four-top quark 
operators by studying the $t \bar{t} b \bar{b}$ 
signature at the LHC. Furthermore, single and pair top quark 
production data is used for finding the bounds on the four-top 
quark coefficients~\cite{Buckley:2015nca,Degrande:2010kt}.
Moreover, four-top quark production has been utilized to probe 
four-top quark operators~\cite{Degrande:2010kt,CMS:1900mtx,Banelli:2020iau}.
The ATLAS collaboration has searched for four-top quark production at $\sqrt{s} = 8$ TeV
to probe four-top quark contact interactions~\cite{Aad:2015gdg,Aad:2015kqa}.
The two top-two light quark and two top-two lepton operators have been studied in top
pair production in the hadron and lepton colliders in Ref.~\cite{AguilarSaavedra:2010zi}.
Also, the effect of four-fermion operators involving only one 
top quark in single top production and top decay width has 
been considered in this paper as well. 
Ref.~\cite{Brivio:2019ius,Hartland:2019bjb} have provided a comprehensive global analysis using the top
quark LHC data to probe four-quark operators including two top-two light quark operator.
The effect of the one top-light quarks operators on the 
polarization of single top quarks in the t-channel process at 
the LHC has been studied in Ref.~\cite{Aguilar-Saavedra:2017nik}. 
Moreover, by using the result of LHC searches for rare 
top decay $t \rightarrow Z j$, bounds on the operators involving one top-one 
light quark-leptons has been determined in Ref.~\cite{Fox:2007in, Durieux:2014xla, Chala:2018agk}.

However, the four-fermion operators with three top and one light quarks 
have not received more attention, since hardly any experimental 
observables are sensitive to these operators~\cite{deBlas:2014mba}. 
In this paper, we explore this kind of four-fermion operators in 
three top quarks and three top quarks+jet productions in hadron colliders.
The cross-section production of the triple top quarks at the hadron 
collider is small compared with other top quark 
production channels in the SM framework~\cite{Barger:2010uw}. 
The reason for this relatively small cross-section is the existence 
of b-quark in the initial state (most of the time) and 
the weak couplings in these processes.  
Since the beyond 
SM models can enhance the rate of triple top quark production, 
this signature can be valuable to show the effect of the new physics.

Previously, the triple top production has been considered for exploring some beyond SM 
models~\cite{Barger:2010uw,Hou:2019gpn,Kohda:2017fkn,Han:2012qu}. 
For example in Ref.~\cite{Malekhosseini:2018fgp,Khanpour:2019qnw}, authors
utilized the three top production 
for constraining the top quark flavor changing neutral couplings 
(FCNC) with the gluon, Higgs and Z boson, and photon.
Moreover, this signature has been considered
in Refs.~\cite{Cao:2019qrb} and \cite{Chen:2014ewl} for probing the contact
interactions at the LHC with 13 and 14 TeV center-of-mass energies.

In this study, we consider a full set of 
four-fermion operators involving three top quarks. 
A detailed analysis is performed for the triple top 
production to probe this kind of contact 
interaction at the high energy LHC (HE-LHC)~\cite{Azzi:2019yne,Cepeda:2019klc} 
and the Future Circular Collider (FCC-hh)~\cite{Benedikt:2018csr}
with integrated luminosity of 15 ab$^{-1}$ and 30 ab$^{-1}$, respectively.
The unique signature of two same-sign leptons final state 
is taken into account in order to suppress the SM backgrounds.
By considering the signal scenario, we consider all the relevant background processes 
which have similar final state topology.
Furthermore, parton showering and hadronization as well as the
realistic detector simulations are taken into consideration. 
Although our aim here is probing the Wilson coefficients of four-fermion operators 
involving three top quarks in triple tops production, finally we constrain the parameters 
of the UV complete model containing a leptophobic gauge boson $Z'$ by using the limits 
on the Wilson coefficients. 

The rest of the paper is organized as follows. 
The theoretical framework considered in this study
is introduced in section~\ref{sec:model}. 
The details of our analysis strategy is presented in section~\ref{sec:Analysis}.
The sensitivity of the future high energy hadron colliers to the Wilson coefficients 
of four-fermion operators involving the three top quarks appear
in section~\ref{sec:sensitivity}. 
Finally, the summary of the paper is presented in section~\ref{sec:summary}.

\section{Theoretical formalism}\label{sec:model}

In this section, we introduce the operators involving three top and a
 light quarks in the context of the SMEFT.
All the new degrees of freedom are integrated out in this framework
and their effects are encoded in the new interaction between the SM 
fields with the appearance of the higher dimensional operators in the Lagrangian. 
These effective operators respect the Lorentz and SM gauge symmetries. 
By considering the baryon and lepton conservation, the SMEFT Lagrangian 
up to the dimension-six operators are written as follows,
\begin{equation}
\mathcal{L}_{\rm{SMEFT}}=\mathcal{L}_{\rm{SM}}+\sum_{i} \frac{C_{i} O_{i}}{\Lambda^{2}},
\end{equation}
where the first term is the SM Lagrangian containing dimension-four operators. 
The dimension-six operators and their corresponding coefficients are shown by $O_{i}$ and $C_{i}$, respectively.
The scale of new physics is indicated by $\Lambda$.
Among the four-fermion contact interactions, following 
operators could generate the interaction terms between three top 
and one light quarks~\cite{Grzadkowski:2010es,AguilarSaavedra:2018nen},
\begin{eqnarray}
	O_{q q}^{1(i j k l)}&=&\left(\bar{q}_{i} \gamma^{\mu} q_{j}\right)\left(\bar{q}_{k} \gamma_{\mu} q_{l}\right), \nonumber \\
	O_{q q}^{3(i j k l)}&=&\left(\bar{q}_{i} \gamma^{\mu} \tau^{I} q_{j}\right)\left(\bar{q}_{k} \gamma_{\mu} \tau^{I} q_{l}\right), \nonumber \\
	O_{u u}^{ (i j k l)}&=&\left(\bar{u}_{i} \gamma^{\mu} u_{j}\right)\left(\bar{u}_{k} \gamma_{\mu} u_{l}\right),\nonumber \\
	O_{q u}^{1(i j k l)}&=&\left(\bar{q}_{i} \gamma^{\mu} q_{j}\right)\left(\bar{u}_{k} \gamma_{\mu} u_{l}\right), \nonumber \\
	O_{q u}^{8(i j k l)}&=&\left(\bar{q}_{i} \gamma^{\mu} T^{A} q_{j}\right)\left(\bar{u}_{k} \gamma_{\mu} T^{A} u_{l}\right), 
\end{eqnarray}
where $q$ and $u$ indicate the left-handed quark doublet and right-handed quark singlet, respectively,
and $i,j,k,l$ are flavour indices. 
The Pauli matrices are shown by $\tau^{I}$ and
$ T^{A}=\lambda^{A}/2$ where $\lambda^{A}$ are Gell-Mann matrices.
The relevant effective Lagrangian, containing interaction between three 
top quarks and one light quark, is obtained as follows,
\begin{eqnarray}
	\mathcal{L}_{\mathrm{eff}}&=&
	 \frac{1}{\Lambda^{2}}[C_{q q_{u}} 
	 (\bar{t}_{L} \gamma^{\mu} t_{L})
	 (\bar{t}_{L} \gamma_{\mu} u_{L})+
	 C_{u u}(\bar{t}_{R} \gamma^{\mu} 
	 t_{R})(\bar{t}_{R} \gamma_{\mu} u_{R}) \nonumber \\
	&+& C_{q u_{_R}} (\bar{t}_{L} 
	\gamma^{\mu} t_{L})(\bar{t}_{R} \gamma_{\mu} u_{R})+
	C_{q u_{_L}}(\bar{t}_{R} \gamma^{\mu} t_{R})(\bar{t}_{L} \gamma_{\mu} u_{L})  \nonumber \\
	&+& C_{q q_{c}} (\bar{t}_{L} \gamma^{\mu} t_{L})
	(\bar{t}_{L} \gamma_{\mu} c_{L})+C_{c c}(\bar{t}_{R} 
	\gamma^{\mu} t_{R})(\bar{t}_{R} \gamma_{\mu} c_{R}) \nonumber \\
	&+& C_{q c_{_R}} (\bar{t}_{L} \gamma^{\mu} t_{L})
	(\bar{t}_{R} \gamma_{\mu} c_{R})+C_{q c_{_L}}(\bar{t}_{R} \gamma^{\mu} t_{R})
	(\bar{t}_{L} \gamma_{\mu} c_{L})]  \nonumber \\
	&+&h.c.
\label{eq:eff-lag}
\end{eqnarray}

In this study, we consider operators containing the $u$-quark and $c$-quark separately. 
In the next section, the above effective Lagrangian is employed to study 
the triple top quark production at the future high energy hadron colliders
to probe the four-fermion operators.
\begin{figure}[t]
\begin{center}
\vspace{0.40cm}
\resizebox{0.42\textwidth}{!}{\includegraphics{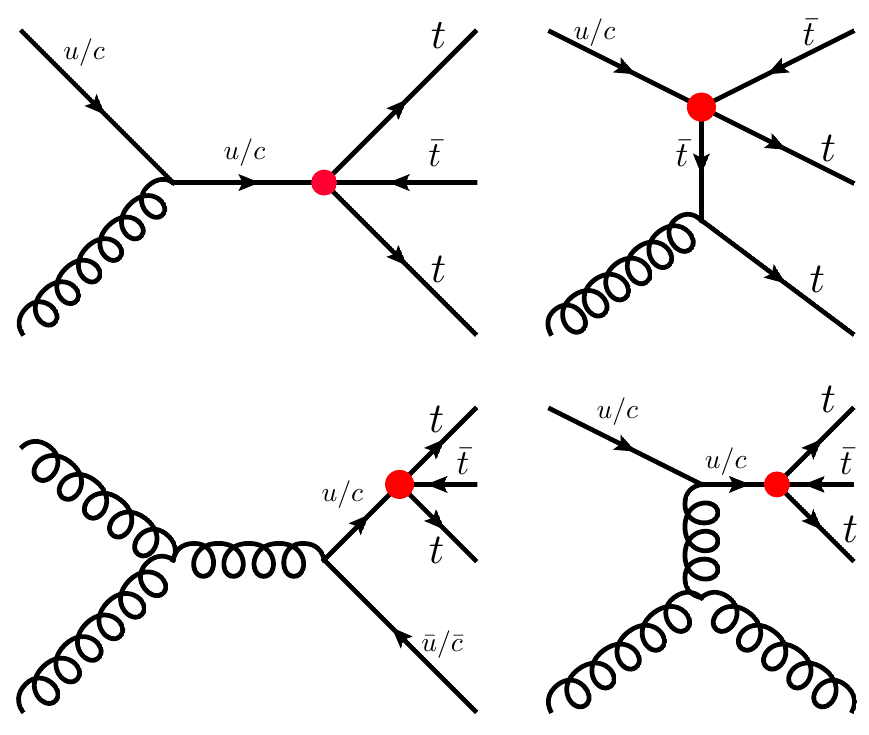}}   	
\end{center}	
\caption{Representative Feynman diagrams for the signal. 
The triple top and triple top in association with 
a jet production in presence of the four-fermion contact 
interactions are shown in the first row and second row, respectively.}
\label{Feynman}
\end{figure}

\begin{figure*}[t]
	\begin{center}
		\resizebox{0.8\textwidth}{!}{\includegraphics{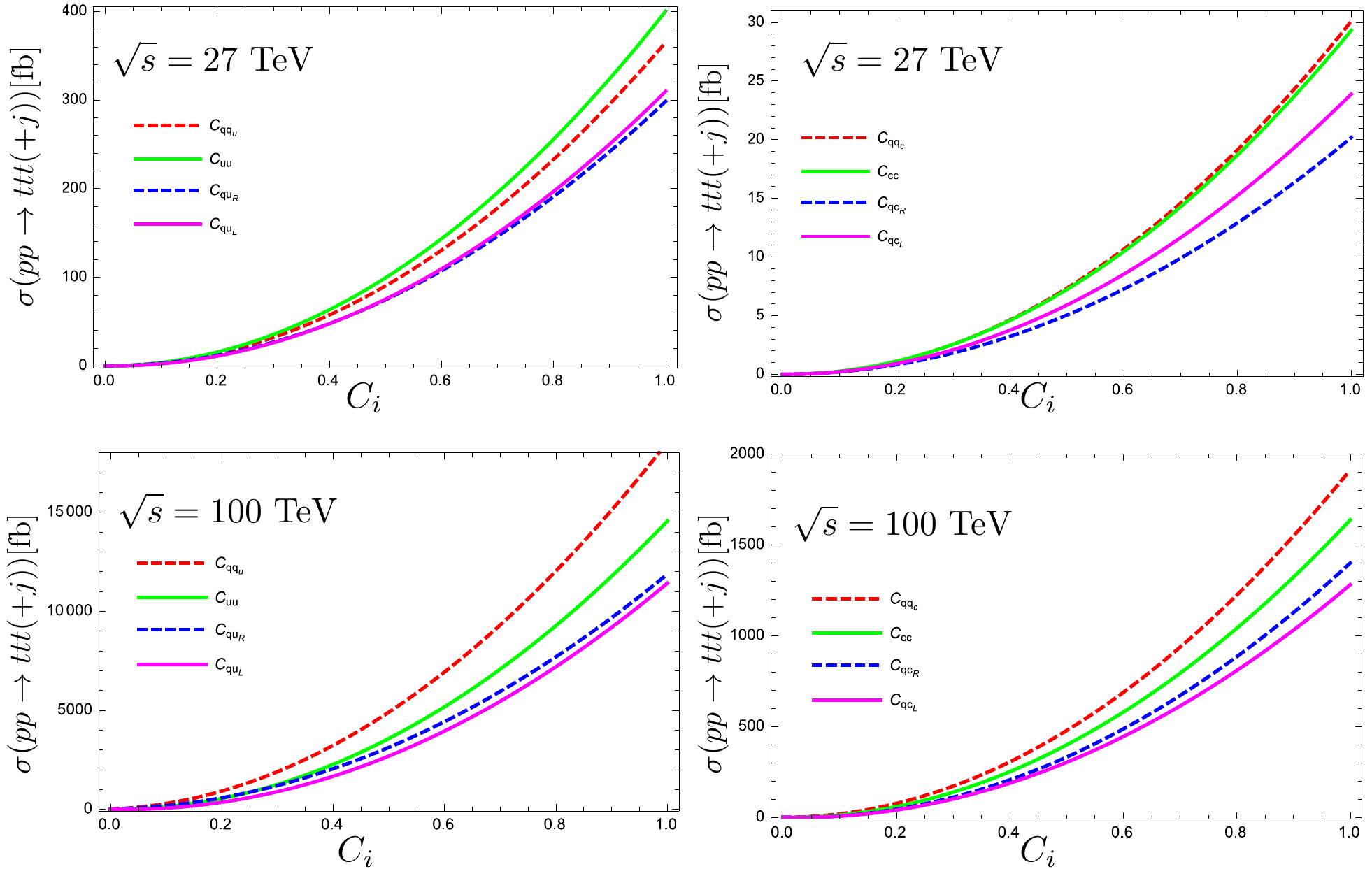}}   	 	\\
			\end{center}	
\caption{The signal cross section  including the branching 
ratios as a function of Wilson coefficients 
for $\Lambda=1$~TeV after applying the preselection at the center of mass energy 27 TeV and 100 TeV. }
\label{Cross-Sction}
\end{figure*}

\section{Analysis Strategy}\label{sec:Analysis}

In this section, we present our signal and the related SM background processes.
We also describe in details the event generation and simulation method.
As mentioned before, we are interested in the triple top quark 
production(+jet) at the future hadron collider to search for 
the four-fermion contact interactions.
We perform the analysis for the HE-LHC and FCC-hh working at the center of
mass energies of $\sqrt {s} = 27~\rm {TeV}$ and $\sqrt {s} = 100~\rm {TeV}$, respectively. 

The signal process is three top quarks production where two same-sign 
top quarks decay leptonically and the other one decay hadronically. 
We also add three top quarks production in association with a jet to the signal process. 
Consequently, the final state consists of two same-sign charged leptons (electron or muon), 
three $b$-jets, two or three light-jets, and missing energy due to the presence of the neutrino:
\begin{eqnarray}
p p &\rightarrow & t t \bar{t} (+j)  \rightarrow \ell^{+} \ell^{+}  
+ 3~\text{b-jet} + 2(+1)~\text{light-jet}+ 
E_T{\hspace{-0.43cm}/}\hspace{0.35cm} , \nonumber \\
p p &\rightarrow & \bar{t} \bar{t} t (+j)  \rightarrow \ell^{-} 
\ell^{-} + 3~\text{b-jet} + 2(+1)~\text{light-jet} +
E_T{\hspace{-0.43cm}/}\hspace{0.35cm} .\nonumber
\end{eqnarray}
Fig.~\ref{Feynman} presents some representative Feynman diagrams of the signal process. 
The triple top and triple top in association with a jet production in presence of 
the four-fermion contact interactions are shown in the first row and second row, respectively.
The effective Lagrangian of Eq.~\eqref{eq:eff-lag} is implemented in 
the \feynrules~package~\cite{Alloul:2013bka} and the obtained 
UFO module~\cite{Degrande:2011ua} is inserted in the \madgraph 5 package~\cite{Alwall:2014hca} 
for generating the signal events.

We generate eight different signal samples corresponding to the eight different 
Wilson coefficients (Eq.~\ref{eq:eff-lag}) to probe each of them separately.
The signal cross section including the branching ratios as a function of different 
Wilson coefficients for the HE-LHC and the FCC-hh colliders at the center of 
mass energy 27 TeV and 100 TeV are depicted in Fig.~\ref{Cross-Sction}.
The plots for Wilson coefficients related to $u$-quark and $c$-quark are shown separately. 
As we expected, the cross section values corresponding to $u$-quark are 
around one order of magnitude larger than the ones related to $c$-quark
due to the large parton distribution function (PDF) 
of the $u$-quark in the proton compared to the $c$-quark. 
In order to generate the signal events, we set $\Lambda=1$~TeV and apply 
preselection cuts, $p_{T} \geq 10$ GeV and $|\eta| < 3$ on all decay products.

\begin{figure*}[t]
	\begin{center}
		\vspace{0.40cm}
		\resizebox{0.4\textwidth}{!}{\includegraphics{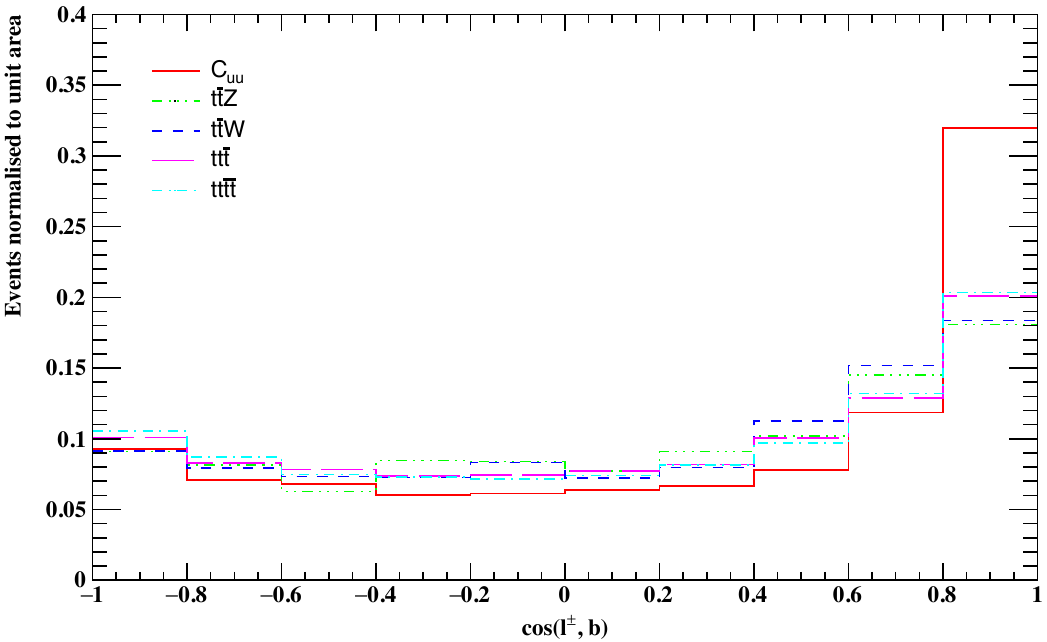}}   		
		\resizebox{0.4\textwidth}{!}{\includegraphics{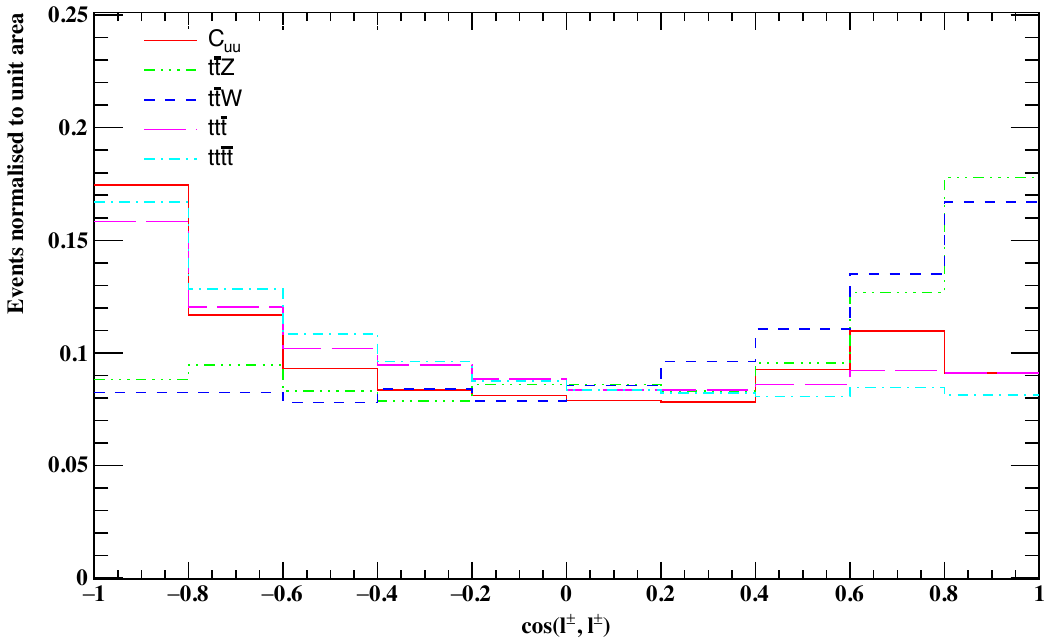}}  \\
		\resizebox{0.4\textwidth}{!}{\includegraphics{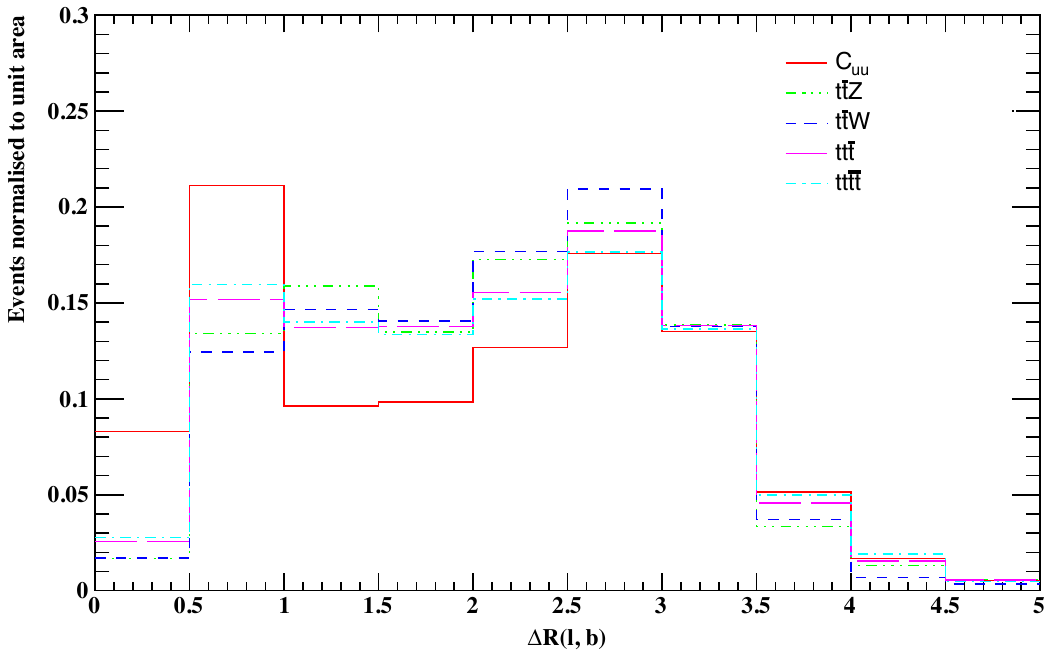}}  
		\resizebox{0.4\textwidth}{!}{\includegraphics{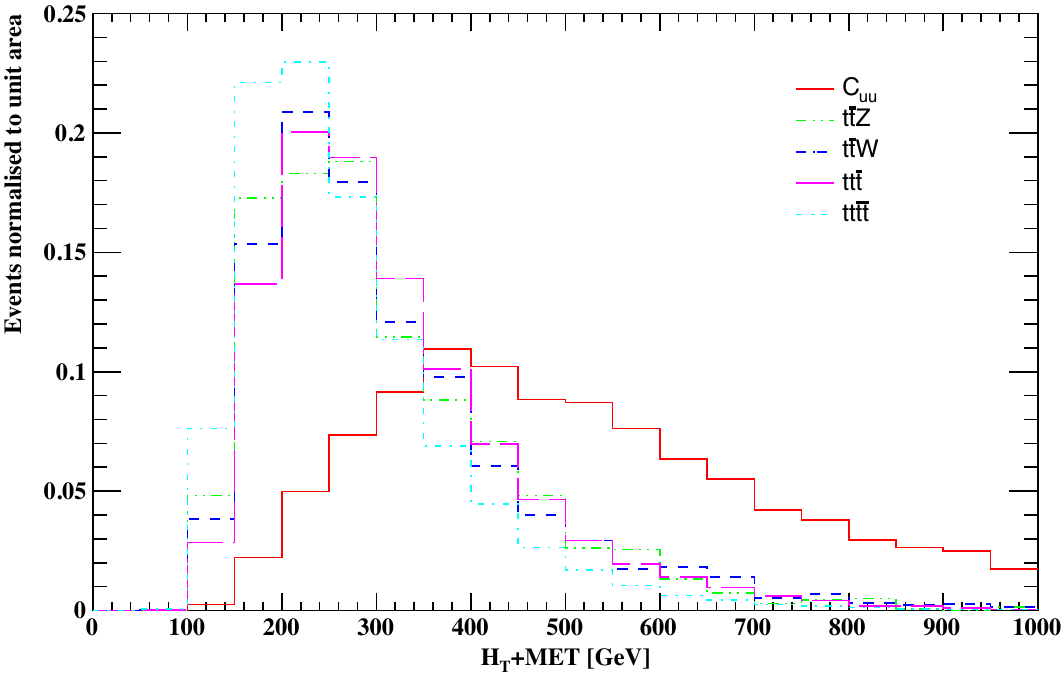}}  \\
		\caption{ The kinematic distributions of the triplet top(+jet) signal (for $C_{uu}$ coupling and $\Lambda=1$~TeV)
and the main SM backgrounds, $t \bar t Z$, $t \bar t W$, SM three top ($t \bar t t$, $\bar t t \bar t$) 
and SM four top production ($t \bar{ t} t \bar{t}$), at the center of mass energy of $\sqrt{s} = 100$ TeV. } 
\label{fig:Distribution}
	\end{center}
\end{figure*}

Considering our signal signature, a complete 
set of the main SM background processes are considered in our study which includes the
$t \bar{t} Z$, $t \bar{t} W^{\pm}$, $W^{+} W^{-} Z$ and four top quark production $t t \bar{t} \bar{t}$.
We also add the SM triple top quark production to the background process. 
However, since the three top quarks cannot generate in the framework 
of the SM alone, we consider $ttt+j$ and $ttt+W$ as the SM triple top quark 
background. All the mentioned SM backgrounds with their decay products are listed below:
\begin{eqnarray}
p p &\rightarrow & t \bar{t} Z \rightarrow 3~\ell + 2~\text{b-jet} + 2~\text{light-jet}+ E_T{\hspace{-0.43cm}/}\hspace{0.35cm} , \nonumber \\
p p &\rightarrow & t \bar{t} W^{\pm} \rightarrow 2~\ell + 2~\text{b-jet} + 2~\text{light-jet}+ E_T{\hspace{-0.43cm}/}\hspace{0.35cm} , \nonumber \\
p p &\rightarrow &  W^{+} W^{-} Z \rightarrow 3~\ell + 2~\text{light-jet}+ E_T{\hspace{-0.43cm}/}\hspace{0.35cm} , \nonumber \\
p p &\rightarrow & t t \bar{t} \bar{t}  \rightarrow 2~\ell + 4~\text{b-jet} + 4~\text{light-jet} +E_T{\hspace{-0.43cm}/}\hspace{0.35cm} ,\nonumber \\
p p &\rightarrow &  t t \bar{t}(t \bar{t} \bar{t} )+ j \rightarrow 2~\ell + 3~\text{b-jet} + 3~\text{light-jet} +E_T{\hspace{-0.43cm}/}\hspace{0.35cm} ,\nonumber\\
p p &\rightarrow &  t t \bar{t}(t \bar{t} \bar{t} ) + W \rightarrow 2 (3)~\ell + 3~\text{b-jet} + 4(2)~\text{light-jet} +E_T{\hspace{-0.43cm}/}\hspace{0.35cm}.\nonumber
\end{eqnarray}

In order to generate the signal and background samples, the \madgraph5 package is used.
The following SM inputs are used in all numerical calculation: $m_{\text{top}}=173.3$ GeV,
$m_{Z}=91.187$ GeV,$\alpha_{ew}=1/127.90$, and $\alpha_s=0.1184$~\cite{Tanabashi:2018oca}. 
We employ the leading order of the {\tt NNPDF23L01} as the parton distribution function~\cite{Ball:2012cx,Ball:2017nwa}.
Also, the dynamical scale is used for renormalization and factorization scale.
Some preselection cuts such as $p_{T} \geq 10$ GeV and $|\eta| < 3$ are applied
on all objects in the final state at the generator level.
The \pythia~package~\cite{Skands:2014pea} is used for parton showering and hadronization. 
To simulate the signal and backgrounds including up to one additional jet in the final state, we employ 
the MLM matching scheme to avoid double counting~\cite{Mangano:2006rw,Frederix:2012ps}.

Furthermore, the \delphes~package~\cite{deFavereau:2013fsa} is utilized to model the detector performance.
We should note here that, for our HE-LHC analysis, we use the \delphes framework to 
perform a comprehensive high luminosity (HL) CMS detector response simulation.
To this end, the HL-LHC detector card configuration available in the \delphes is 
used which includes the high configuration of the CMS detector~\cite{Azzi:2019yne,Cerri:2018ypt}. 
For the case of FCC-hh projections, we use the default FCC detector card configuration 
implemented in \delphes~\cite{Oyulmaz:2019jqr}.
Considering these configurations, the efficiency of $b$- and $c$-tagging and light-flavor quarks
misidentifications rates are assumed to be the jet transverse 
momentum dependent ($p_T$)~\cite{Khanpour:2019qnw}.

\begin{table*}[tbh]
\begin{center}
\begin{tabular}{ c | c c c c c c c c  }
\hline \hline 
$\sqrt {s}$ = 27~TeV (HE-LHC)      ~&~  $C_{q q_{u}}$  ~&~ $C_{u u}$   ~&~ $C_{q u_{_R}}$  ~&~ $C_{q u_{_L}}$ ~&~ $C_{q q_{c}}$  ~&~  $C_{c c}$ ~&~  $C_{q c_{_R}}$  ~&~ $C_{q c_{_L}}$   \\   \hline  
Preselection cuts & 373.44 & 392.83 & 298.17 & 310.94  & 30.52   & 30.01  & 20.55 & 24.15   \\ 
Cut (I)            & 75.18  & 86.89 &  75.80 &  79.43  & 7.61    & 8.92   & 6.08  & 7.11    \\
Cut (II)           & 74.81  & 86.42 &  75.38 &  79.06  & 7.57    & 8.85   & 6.04  & 7.07    \\
Cut (III)          & 49.43  & 51.15 &  48.16 &   51.75 &  4.84   &  5.13  & 3.75  & 4.54    \\	
Cut (IV)           & 19.25  & 16.96 &  18.97 &   19.13  &  1.94  &  1.71  & 1.45  & 1.67    \\ 	
Cut (V)            & 11.03  & 12.48 &  11.10 &   11.00  &  0.96  &  1.11  & 0.77  & 0.84    \\ 	                                                        
\end{tabular}
\begin{tabular}{ c | c c c c c c c c}
\hline \hline   
$\sqrt {s}$ = 100~TeV (FCC-hh)      ~&~  $C_{q q_{u}}$  ~&~ $C_{u u}$   ~&~ $C_{q u_{_R}}$  ~&~ $C_{q u_{_L}}$ ~&~ $C_{q q_{c}}$  ~&~  $C_{c c}$ ~&~  $C_{q c_{_R}}$  ~&~ $C_{q c_{_L}}$   \\   \hline  
Preselection cuts &  17862.71  & 14379.80 & 12376.94 &  12957.84 & 1881.59 & 1649.85  & 1431.31  &  1403.15   \\ 
Cut (I)           & 2147.37    & 1830.48   & 2322.97   &  2442.03 & 382.37   & 388.14  & 391.41   &  383.42   \\
Cut (II)          & 2138.79    & 1820.91   & 2312.63   &  2432.19 & 380.58   & 386.02  & 389.47   &  381.53    \\
Cut (III)         & 1768.59    & 1420.94  & 1882.59   &  1993.76 & 316.52  & 303.08   & 318.43   &  314.05    \\	
Cut (IV)          & 875.00 	   & 602.58   & 937.98    &  954.73  & 160.01  & 132.25   & 161.49   &  151.48   \\ 	
Cut (V)           & 567.05     & 481.43   & 616.99    &  611.22  & 95.39   & 97.98    & 98.58    &  91.65   \\  \hline \hline                                                           
\end{tabular}
\end{center}
\caption{ The values for $\sigma_{0}$ (in the unit of fbTeV$^4$) including the branching ratios 
for eight signal scenarios, before and after the selection criteria for the HE-LHC and FCC-hh colliders.}
\label{tab:Cross-Signal}
\end{table*}

\begin{table*}[tbh]
\begin{center}
\begin{tabular}{ c | c c c c c }
\hline  \hline  
$\sqrt {s}$ = 27~TeV (HE-LHC)      ~&~  $t \bar t Z$  ~&~ $t \bar t W$   ~&~ $W W Z$  ~&~ SM ($t \bar t t$, $\bar t t \bar t$) ~&~  SM ($t \bar t  t   \bar t$) \\   \hline  
Preselection cuts           ~&~      33.92      ~&~  17.45    ~&~    3.03   ~&~  0.188  ~&~  0.940 \\ 
Cut (I)    ~&~ 4.26  ~&~ 6.57 ~&~  0.36    ~&~  0.032  ~&~ 0.271 \\
Cut (II)   ~&~ 3.62  ~&~ 5.90  ~&~  0.28    &  0.029    ~&~ 0.251  \\
Cut (III)  ~&~ 1.31   ~&~  1.06  ~&~   0.011     ~&~   0.022 ~&~  0.227   \\	
Cut (IV)   ~&~ 0.132  ~&~ 0.117 ~&~  0.00003    ~&~   0.009 ~&~  0.121 \\ 	
Cut (V)    ~&~ 0.02  ~&~ 0.02 ~&~  0.0   ~&~   0.002  ~&~  0.033 \\	                                                         
\end{tabular}
\begin{tabular}{ c | c c c c c }
\hline  \hline  
$\sqrt {s}$ = 100~TeV (FCC-hh)      ~&~  $t \bar t Z$  ~&~ $t \bar t W$   ~&~ $W W Z$  ~&~ SM ($t \bar t t$, $\bar t t \bar t$) ~&~  SM ($t \bar t  t   \bar t$) \\   \hline  
Preselection cuts    ~&~      343.44      ~&~   73.16    ~&~     12.45   ~&~  2.83  ~&~   15.18    \\ 
Cut (I)   ~&~ 39.424   ~&~ 32.09 ~&~  1.30  ~&~  0.54 ~&~  5.03   \\
Cut (II)  ~&~ 33.23    ~&~ 28.96  ~&~  1.01   ~&~  0.50 ~&~  4.68  \\
Cut (III) ~&~  19.59  ~&~  11.53  ~&~   0.11    ~&~   0.46   ~&~  4.60   \\	
Cut (IV) ~&~ 2.76 	~&~1.53 ~&~  0.00056   ~&~  0.25   ~&~ 3.27   \\    
Cut (V)   ~&~ 0.728 	~&~   0.310   ~&~    0.0   ~&~   0.088     ~&~ 0.717  \\    \hline    \hline                                              
\end{tabular}
\end{center}
\caption{ Cross section (in the unit of fb)  including the branching ratios for the 
$t \bar t Z$, $t \bar t W$, $WWZ$, SM three top ($t \bar t t$, $\bar t t \bar t$) 
and SM four top production ($t \bar t  t   \bar t$), 
before and after the selection criteria, for the HE-LHC and FCC-hh colliders.}
\label{tab:Cross-Back}
\end{table*}

For selecting the event rate for all signal scenarios, we require to apply the following selection cuts:
\begin{itemize}
\item Cut (I): $n^{\ell}=2\ell^{\pm \pm}$, $|\eta^{\ell}|<2.5$, $p_T^{\ell} > 10$ GeV, $M_{\ell^\pm \ell^\pm} > 10$ GeV.
\item Cut (II): $E_T{\hspace{-0.43cm}/}\hspace{0.35cm} >30$ GeV.
\item Cut (III): $n^{\rm jets} \geq 5$  jets, $|\eta^{\rm jets}|<2.5$, $p_T^{\rm jets}>20$ GeV, $\Delta R(\ell, j_i) \geq 0.4$, $\Delta R(j_i, j_j) \geq 0.4$.
\item Cut (IV): $n^{\rm b-jets}  \geq 3 \, \, \rm{b-jets}$\,,
\end{itemize}
where $\Delta R =\sqrt{(\Delta \phi)^{2}+(\Delta \eta)^{2}}$.
In order to suppress the contributions of SM backgrounds, 
we look at the different kinematic distributions to find a proper secondary cut.  
In Fig.~\ref{fig:Distribution}, we present some selected distributions for the signal sample and all 
main background processes at the center of mass energy of 100 TeV. 
The signal sample is generated for the $C_{u u}=1$ and $\Lambda=1$~TeV. 
The distribution of cosine between the leading charged-lepton ($\ell^\pm$) and the leading $b$-jet, $\cos(\ell, b)$, and 
the distribution of cosine between two same-sign charged leptons, $\cos(\ell^\pm, \ell^\pm)$, are illustrated.
The angular distance between the leading charged-lepton and the leading $b$-jet $\Delta R (\ell, b)$ and
$H_T + E_T{\hspace{-0.43cm}/}\hspace{0.35cm}$ distribution are shown as well. 
Here, the definition of the $H_T$ is the sum of all the leptons and jets $p_T$.

All the distributions are presented after applying all the selection cuts described above.
The cut efficiency of $W^{+} W^{-} Z$ background is around $10^{-5}$, therefore the
distribution of this background isn't shown in Fig.~\ref{fig:Distribution}.
From $H_T + E_T{\hspace{-0.43cm}/}\hspace{0.35cm}$ distribution, 
it is clear that the signal has more spread distribution which extends up to around $1$~TeV. 
The $H_T + E_T{\hspace{-0.43cm}/}\hspace{0.35cm}$ distribution 
for all other signal scenario have the same behavior. 
In order to separate more signal from the background events, 
we could add another criterion to our selection cuts as follows, 
\begin{itemize}
\item Cut (V): $H_T + E_T{\hspace{-0.43cm}/}\hspace{0.35cm}  \geq 340$ GeV.
\end{itemize}

It can be seen from Fig.~\ref{Cross-Sction}, for the different signal scenarios, 
we can write the dependence of the cross sections to the operator coefficients as follows,
\begin{equation}
\sigma = \sigma^i_0\times (\frac{C_{i}}{\Lambda^2})^2.
\label{Eq:sigma}
\end{equation}
In Table~\ref{tab:Cross-Signal}, these values of $\sigma^i_0$ for eight signal scenarios (different $C_i$s) 
before and after the selection criteria are presented for the HE-LHC and FCC-hh colliders 
at the center of mass energy 27 TeV and 100 TeV. 
Moreover, the expected cross sections include the branching ratios
for the relevant SM backgrounds are shown in Table~\ref{tab:Cross-Back} which are reported in the unit of fb. 
We should highlight here that the values in the first rows of two tables are reported 
after applying the preselection cuts.

It should be mentioned here that, we consider two other potential sources of background categories. 
The first category rises up from mismeasurement of the charge of leptons. 
The same-sign signature can be appeared in processes like $t\bar{t}$ dilepton channel, $tW$ dilepton channel, 
and Drell-Yan, if the charge of a lepton is mismeasured and they can play the role of the background process. 
The efficiencies for these background processes are completely negligible after applying the secondary cuts,
particularly when we require the cuts on the number of jets ($n^{\rm jets} \geq 5$). 
The efficiency of whole background is less than $0.01\%(0.1\%)$ at center of mass energy of $27(100)$ TeV.
Also, these efficiencies should multiply in the charge mismeasurement probability 
which is about $3.3 \pm 0.2 \times 10^{-4}$ for electron, whereas this value for muon is tiny~\cite{Chatrchyan:2012fla}.
As a result, the contributions from these backgrounds are rather small and safely could be neglected.

The second category comes from when a jet is misidentified as a lepton.
Some example of this sort of background process are: (i)  $W +$ jets events in which $W$ decays leptonically 
and one of the jets is misidentified as a lepton; (ii) Semi-leptonic $t\bar{t}$ events in which the second leptons
originated from misidentification of a jet; and (iii) the t-channel single top production 
in which top quark decay leptonically and one of the jets fakes the lepton. 
We found pretty small efficiencies for these backgrounds after imposing the secondary cuts. Hence 
we neglect the contributions of these two sources of backgrounds in our analysis.

\begin{table*}[tbh]
\begin{center}
\begin{tabular}{|c|c|c|c|c|}
\hline \multirow{2}{*} { Wilson Coefficient} & \multicolumn{2}{|c|} { HE-LHC, $15~\mathrm{ab}^{-1}$} & \multicolumn{2}{c|} { $\mathrm{FCC}-\mathrm{hh}, 30~\mathrm{ab}^{-1}$}  \\
\cline { 2 - 5 } & $\delta=0\%$ & $\delta=10 \%$ & $\delta=0\%$ & $\delta=10 \%$  \\
\hline \hline 
$C_{q q_{u}}$  & $[\, -0.042, \, 0.011\, ]$ & $[\, -0.044, \, 0.013\, ]$ & $[\, -0.018, \, 0.0017 \, ]$ &$[\, -0.019, \, 0.0019 \, ]$  \\
$C_{u u}$  & $[\, -0.008, \, 0.047 \, ]$ & $[\, -0.009, \, 0.048 \, ]$ &  $[\, -0.055, \, 0.0007\, ]$ & $[\, -0.056, \, 0.0008\, ]$  \\
$C_{q u_{_R}}$ &  $[\, -0.020, \, 0.023 \, ]$&  $[\, -0.021, \, 0.025 \, ]$ &$[\, -0.003, \, 0.013 \, ] $ & $[\, -0.003, \, 0.014 \, ] $\\
$C_{q u_{_L}}$ & $[\, -0.021, \, 0.022 \, ]$ & $[\, -0.023, \, 0.024 \, ]$& $[\, -0.002, \, 0.011 \, ]$ & $[\, -0.002, \, 0.012 \, ]$  \\
\hline
 $C_{q q_{c}}$  & $[\, -0.068, \, 0.078\, ]$ & $[\, -0.074, \, 0.084\, ]$ & $[\, -0.006, \, 0.028\, ] $& $[\, -0.007, \, 0.029\, ] $\\
$C_{c c}$  &  $[\, -0.067, \, 0.069 \, ]$ & $[\, -0.073, \, 0.075 \, ]$& $[\, -0.007, \,0.023 \, ]$ & $[\, -0.008, \,0.024 \, ]$  \\
$C_{q c_{_R}}$ & $[\, -0.125, \, 0.057 \, ]$ &$[\, -0.131, \, 0.064 \, ]$ & $[\,-0.003, \, 0.044 \, ] $& $[\,-0.004, \, 0.045 \, ] $ \\
$C_{q c_{_L}}$ & $[\, -0.059, \, 0.100\, ]$ & $[\, -0.065, \, 0.106\, ]$ & $[\, -0.003, \, 0.059\, ] $ &$[\, -0.003, \, 0.060\, ] $ \\
\hline
\end{tabular}
\caption{Constraints at 95\% CL on the Wilson coefficients ($C_{i}(1~\rm{TeV})^2/\Lambda^2$) of four-fermion operators 
 at the center of mass energy of 27 TeV (HE-LHC) and 100 TeV (FCC-hh)
 with the integrated luminosity of 15~ab$^{-1}$ and 30~ab$^{-1}$, respectively . 
We consider an overall systematic uncertainty of $0\%$ and $10\%$ on the signal efficiency and number of backgrounds
 At a time one of the coupling is considered in the analysis.}
\label{tab:results} 
\end{center}
\end{table*}

\section{Sensitivity Estimation} \label{sec:sensitivity}                                            %

This section presents the potential sensitivity 
of HE-LHC and FCC-hh colliders to probe the contact interactions.
We demonstrate the upper limits on the Wilson coefficients of the four-fermion interactions involving three top quarks
at $95\%$ confidence level (CL) by analyzing triple top quarks(+jet) production.
A Bayesian approach with a flat prior distribution is employed to estimate the upper limits. 
The probability of observing number of events is assumed to have a Poisson distribution~\cite{Tanabashi:2018oca},

\begin{equation}
L\left(
n_{\text{obs}}, 
n_{\text{S}}, 
n_{\text{B}} 
\right)=
\frac{\left(n_{\text{S}}+
n_{\text{B}}\right)^
{n_{\text{obs}}}}
{n_{\text{obs}} !} 
e^{-\left(n_{\text{S}}+
n_{\text{B}}
\right)},
\end{equation}
here $n_{\text{S}} = \epsilon_{\text{S}} 
\times \cL \times \sigma_{\text{S}}$
where $\epsilon_{\text{S}}$ and $\cL$ are the efficiency of the signal 
and the integrated luminosity, respectively.
The number of background events are given by
$n_{\mathrm{B}} = \epsilon_{\mathrm{B}} 
\times \cL \times \sigma_{\mathrm{B}}$ that
$\epsilon_{\mathrm{B}}$ is the efficiency of the backgrounds.
The upper limit at $95\%$ CL on the number of signal events can be obtained using following formula:

\begin{equation}
\frac{95}{100}=
\frac{\int_{0}^
{n_{\text {limit }}}
L\left(n_{\text{obs}}, 
n_{\text{S}}, n_{\text{B}} 
\right) \text{d} 
n_{\text{S}}}{\int_{0}^{\infty} 
L\left(n_{\text{obs}}, 
n_{\text{S}}, n_{\text{B}} \right) 
\text{d} n_{\text{S}}}.
\end{equation}

In order to obtain the upper limit on the number of signal events, 
we assume that the observed number of events are in consistent with the expected number of 
background events. Then we translate this upper limit to the signal cross section and Wilson coefficients
of four-fermion interaction.
Table~\ref{tab:results} presents the upper limits at 95\% CL on the Wilson coefficients 
of four-fermion operators involving three top quarks
at the center of mass energy 27 TeV and 100 TeV with the integrated luminosity of 15~ab$^{-1}$ and 30~ab$^{-1}$,
respectively. It worth mentioning that, at a time one of the couplings is considered in the analysis.

For finding the upper limits, we scale the production cross section of our main backgrounds
to their next-to-leading (NLO) values.
The NLO corrections to the $t \bar{t} Z$, $t \bar{t} W$ and $t\bar{t}t\bar{t}$ production 
can be found in Ref.~\cite{Azzi:2019yne} for the HE-LHC and Ref.~\cite{Mangano:2016jyj,Maltoni:2015ena} for FCC-hh.
The cross sections for $t \bar{t} Z$, $t \bar{t} W$ and $t\bar{t}t\bar{t}$ are multiplied by the NLO K-factors of
1.2 (1.17), 1.6 (2.2) and 1.3 (1.2) at the HE-LHC (FCC-hh), respectively. 

\begin{figure}
\begin{center}
\vspace{0.40cm}
\resizebox{0.5\textwidth}{!}{\includegraphics{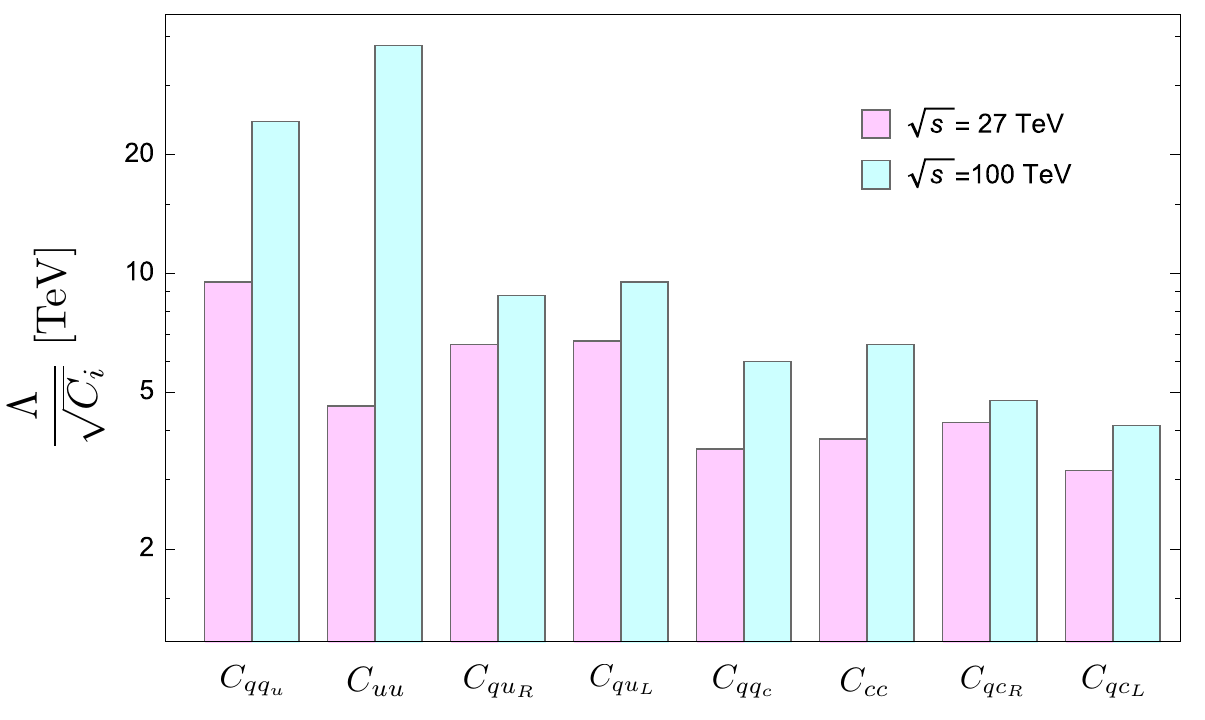}}   	
\end{center}	
\caption{The limits at 95\% CL on $\Lambda/\sqrt{C_{i}}$ 
at the center of mass energy of 27 TeV (HE-LHC) and 100 TeV (FCC-hh)
with the integrated luminosity of 15~ab$^{-1}$ and 30~ab$^{-1}$, respectively .}
\label{Lambda}
\end{figure}

Furthermore for finding more realistic upper limits, the systematic uncertainties should be taken into account.
The source of systematic uncertainties comes from factorization and renormalization scales, 
proton parton distribution function, top quark mass, luminosity measurements, etc. In this analysis, 
we consider an overall systematic uncertainty of $10\%$ on the signal efficiency and the number of backgrounds.
The obtained limits with considering 10\% systematic error are shown in Table~\ref{tab:results} as well. 
As it is clear from the table by considering the 10\% systematic error, we have the weaker upper limits on the Wilson
coefficients of effective operators. 
In Fig.~\ref{Lambda}, we present our bound at 95\% CL on the $\Lambda/\sqrt{C_{i}}$  at HE-LHC and FCC-hh 
with the integrated luminosity of 15~ab$^{-1}$ and 30~ab$^{-1}$, respectively. 

\begin{figure}
\begin{center}
\resizebox{0.4\textwidth}{!}{\includegraphics{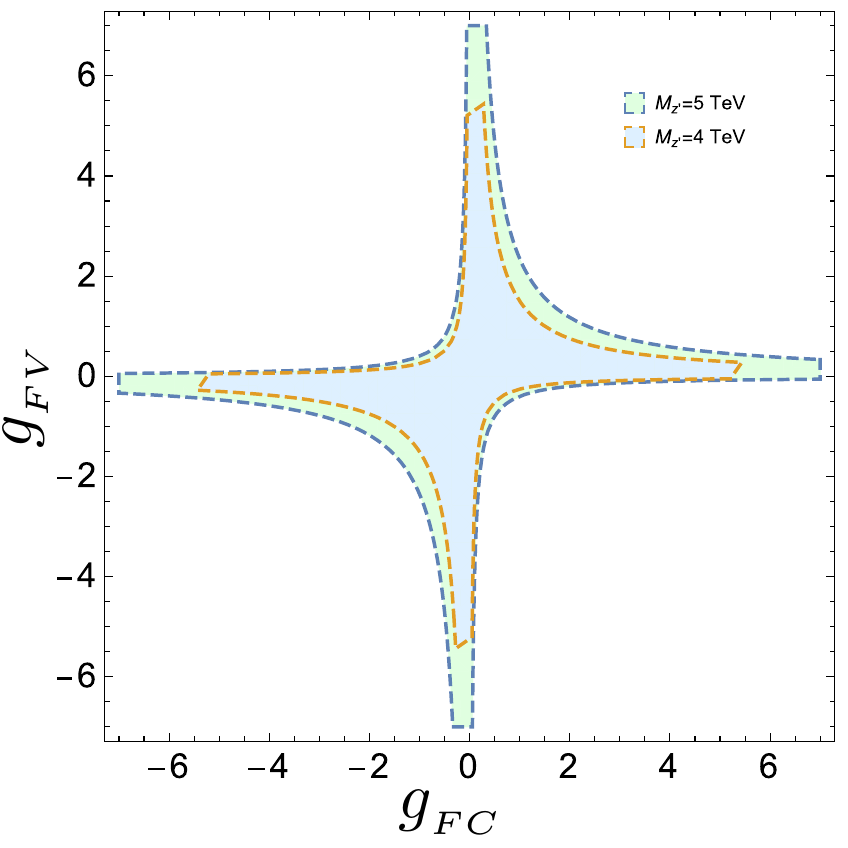}}   	
\end{center}	
\caption{The allowed region (at $95\%$ CL) for $Z'$ parameter space at the center of mass energy 27 TeV 
with the integrated luminosity of 15~ab$^{-1}$ for $M_{Z'}=4$ TeV and $M_{Z'}=5$ TeV. }
\label{fig:LimitZ}
\end{figure}

As mentioned before, one can translate the upper limits on the Wilson coefficients to the parameters of the UV complete 
model. Four instance, a model containing a new leptophobic gauge boson $Z'$ associated with an abelian 
gauge symmetry U(1) with following interaction term can capture some 
of our contact operators at low energy~\cite{Jung:2009jz},
\begin{equation}
\mathcal{L}^{Z'} \supset g_{_{FC}}\bar{ t}_{R} \gamma^{\mu}t_{R} Z'_{\mu}+ \lbrace g_{_{FV}} \bar{u}_{R}\gamma^{\mu}t_{R} Z'_{\mu}+h.c. \rbrace,
\end{equation}
where $g_{_{FC}}$ and $g_{_{FV}}$ are flavour conserving and flavour violating couplings, respectively.
After  integrating the $Z'$ out the obtained effective Lagrangian included flowing term:
\begin{equation}
\mathcal{L}^{Z'}_{\rm{eff}} \supset \frac{-g_{_{FC}} g_{_{FV}} }{2 M^2_{Z'}}(\bar{ t}_{R} \gamma^{\mu}t_{R}) (\bar{u}_{R}\gamma^{\mu}t_{R})+h.c. 
\label{eq:Z}
\end{equation}
By comparing the effective Lagrangians in Eq.~\eqref{eq:eff-lag} and Eq.~\eqref{eq:Z}, 
it is found that $C_{u u}=-g_{_{FC}} g_{_{FV}}/2$ and $\Lambda=M_{Z'}$ at the tree level. 
As a result, we can find the allowed region for $g_{_{FC}}$ and $g_{_{FV}}$ couplings 
by using the upper limits on $C_{u u}$. Fig.~\ref{fig:LimitZ} shows the allowed region for the $Z'$ model 
at the center of mass energy 27 TeV with the integrated luminosity of 15~ab$^{-1}$.
Since we need to be sure about the validity of the EFT approach, 
the assumed $Z'$  masses should be 
sufficiently larger than the relevant energies and momenta of our processes.
As result, we choose the $M_{Z'}=4$ TeV and $M_{Z'}=5$ TeV benchmarks to find the limits on $Z'$ couplings. 
If consider the same value for the flavour conserving and flavour violating couplings ($g=g_{_{FC}}=g_{_{FV}}$), 
the upper limits at $95\%$ CL is found to be $|g|<1.2~(1.5)$ for $M_{Z'}=4~(5)$ TeV.

\section{Summary} \label{sec:summary}                                       %

The Standard Model Effective Field Theory (SMEFT) is a model-independent 
approach to search for new physics in the colliders.
The SMEFT Lagrangian consists of higher-order operators only containing 
the SM degrees of freedom and respecting the Lorentz and the SM gauge symmetry. 
Among leading corrections to the SM Lagrangian, the four-fermion interactions can 
be originated at the tree level in the UV completion model 
therefore they can have large Wilson coefficients.

In this paper, we examine the sensitivity potential of the future high-energy proton-proton
colliders to probe the four-fermion contact interaction involving three top quarks. 
We study the three top quarks and three top quarks+jet productions in the future hadron colliders to explore
the three top-one light quark operators. 
A detailed analysis considering the full set of background 
processes and the real detector simulation is performed. 
We find the upper limits at the $95\%$ CL on the Wilson 
coefficients for the high energy LHC (HE-LHC)
and the Future Circular Collider (FCC-hh) colliders at 
the center of mass energy of 27 TeV and
100 TeV with the integrated luminosity of 15~ab$^{-1}$ and 30~ab$^{-1}$, respectively. 
Furthermore, we consider a UV completion model, a leptophobic $Z'$ boson associated with an abelian 
gauge symmetry U(1) and by matching we find the
model's allowed region for different $Z'$ mass values.

\section*{Acknowledgments} \label{sec:ack}                                            %

Authors are thankful to Mojtaba Mohammadi Najafabadi for valuable comments on the
manuscript and many helpful discussions. SK 
thanks also Mikael Chala for useful discussions.
HK thanks School of Particles and Accelerators, Institute for Research in 
Fundamental Sciences (IPM) and University of Science and Technology of 
Mazandaran for financial support of this research.

\bibliographystyle{JHEP}                                                                                      %

\bibliography{tripletopRef}

\end{document}